\documentclass[a4paper]{article}

\usepackage{ISCSLP2024}
\usepackage{multirow}
\usepackage{amssymb}
\usepackage{pifont}
\usepackage{cite}
\usepackage{comment}
\usepackage{subcaption}
\usepackage{url}

\title{An Investigation of Reprogramming for Cross-Language Adaptation in Speaker Verification Systems}
\name{Jingyu Li$^{*}$, Aemon Yat Fei Chiu$^{*}$, Tan Lee}
\address{Department of Electronic Engineering, The Chinese University of Hong Kong, Hong Kong}
\email{lijingyu0125@link.cuhk.edu.hk, aemon.yf.chiu@link.cuhk.edu.hk, tanlee@ee.cuhk.edu.hk}

\begin{document}

\maketitle
\def\thefootnote{*}\footnotetext{These authors contributed equally to this work.}
\begin{abstract}
Language mismatch is among the most common and challenging domain mismatches in deploying speaker verification (SV) systems. Adversarial reprogramming has shown promising results in cross-language adaptation for SV. Reprogramming is implemented by padding learnable parameters on the two sides of input speech signals. In this paper, we investigate the relationship between the number of padded parameters and the performance of the reprogrammed models. Sufficient experiments are conducted with different scales of SV models and datasets. The results demonstrate that reprogramming consistently improves the performance of cross-language SV, while the improvement is saturated or degraded when using larger padding lengths. The performance is mainly determined by the capacity of the original SV models instead of the number of padded parameters. The SV models with larger scales have higher upper bounds in performance and can endure longer padding without performance degradation.
\end{abstract}

\section{Introduction}

Deep neural network (DNN) models have been adapted widely in speaker verification (SV) systems with success \cite{snyder2017deep,snyder2018x,desplanques20_interspeech}. These systems often face performance degradation in domain mismatch scenarios, for instance, when the input language differs from the training data \cite{940862,1327105,7078603,7919004,8282182,chen20n_interspeech}. Various domain adaptation techniques are being attempted to address this problem. Supervised fine-tuning is a straightforward approach but computationally heavy and overfitting-prone \cite{pmlr-v162-ju22a,9889705}. In \cite{wang2018unsupervised}, adversarial training separated language-specific information from speaker embeddings. This method can be inefficient due to the need for full model parameter adjustments. \cite{li2022coral++,li22m_interspeech} observed that aligning the statistical properties of speaker embeddings across languages improves SV performance in a new language efficiently without modifying the DNN model. However, these adaptation methods lack the learning ability to capture new knowledge outside the extracted embeddings to improve the performance.


Adversarial reprogramming was first proposed in \cite{elsayed2018adversarial}. Learnable parameters are added at the input level to subsequently change the model output for different tasks without modifying the original pre-trained models. \cite{li2023efficient} successfully adopted reprogramming for cross-language adaptation in SV, in which learnable parameters are padded on the beginning and end of the input waveform to change its output speaker embedding. As the pre-trained SV models are frozen during reprogramming, the information from the new language is supposed to be learned from the padded parameters. Thus, a question is raised: does a larger number of padded learnable parameters improve the reprogrammed models' performance? In the present study, we first revisit the use of the reprogramming method in SV and identify its limitations. A modified input padding method is proposed to fit the SV task. Then, we investigate the relationship between the number of padded parameters and the performance of the reprogrammed models with sufficient experiments. The experiments are conducted on three SV models and different scales of datasets. The results indicate that the performance is boosted by reprogramming effectively. Still, the primary factor in determining the upper bound of the reprogrammed results is the capacity of the original SV models. In other words, increasing the padding learnable parameters does not persistently improve the performance.

\section{Reprogramming for Speaker Verification}
\label{ssec:reprogramming}
The principle of the reprogramming method is to modify the DNN model output by altering its input without changing the pre-trained model\cite{elsayed2018adversarial}, e.g., padding learnable parameters to the model input. The typical input modification approach for audio data is appending learnable parameters to the input waveform\cite{yang2021voice2series, li2023efficient}.
Following the settings in \cite{li2023efficient}, the learnable parameters $W=[w_1,w_2, ..., w_n]$ are concatenated on both sides of the original speech waveform $\textbf{\emph{x}}=[x_1,x_2, ..., x_l]$ directly as:
\begin{equation}
    \Tilde{\textbf{\emph{x}}}=[w_1, ...w_{n//2}, x_1, ..., x_l, w_{n//2+1}, ..., w_n]
  \label{eq:concat}
\end{equation}
Denote the speaker embedding extraction model as $F$, and the reprogrammed embedding is represented as $\Tilde{s}=F(\Tilde{\textbf{\emph{x}}})$. A classifier is added after $F$ to predict the input speaker identities using $\Tilde{s}$ as input. During training, only the following modules are trained to decrease the classification loss: learnable parameters $W$ and the back-end classifier.



In the vanilla reprogramming algorithm, the optimization of $W$ requires knowing the structure of $F$ for gradient calculation and back-propagation, which is not available for black-box SV systems. \cite{li2023efficient} proposed a modified reprogramming method for cross-language domain adaptation in SV models, which is suitable for black-box models. Specifically, a lightweight parallel neural network is used to estimate gradients, bypassing the need for full back-propagation through the pre-trained model $F$. It has shown that both vanilla and modified reprogramming achieve significant performance in SV adaptation to a new language.

\begin{figure}[t]
  \centering
  \scalebox{0.95}{
  \includegraphics[width=\linewidth]{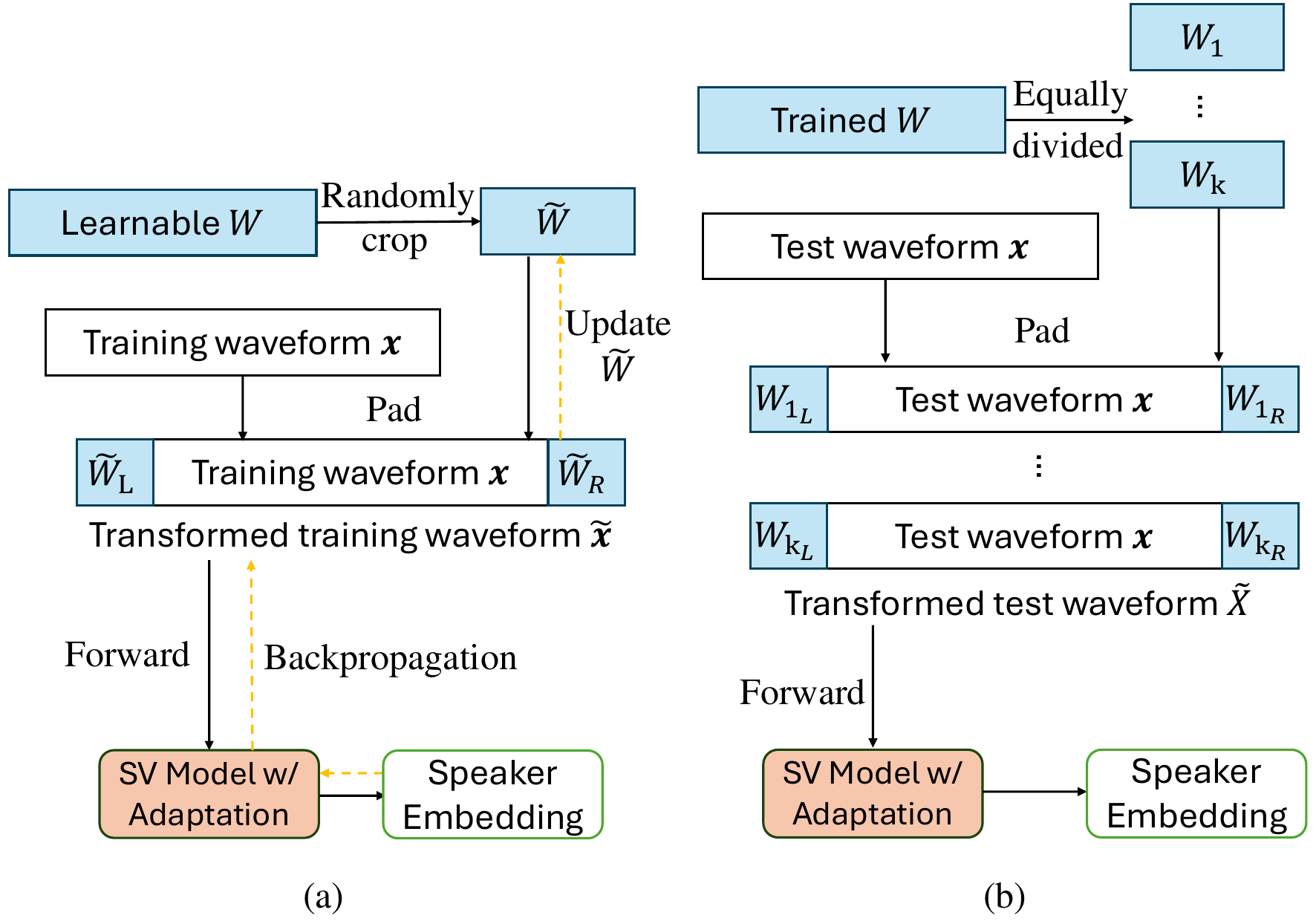}
  }
  \caption{The proposed representation pipeline for (a) Training, (b) Inference}
  \label{fig:proposeinput}
  \vspace{-3mm}
\end{figure}

\section{Methodology}
\subsection{Limitations in current methods}
\label{ssec:limitation}

Let $l$ denote the number of learnable parameters $W$ padded to the input waveform $\textbf{\emph{x}}$. The padded waveform is denoted as $\Tilde{\textbf{\emph{x}}}$. Larger $l$ brings greater learning ability for the model to learn new information from the target domain. In \cite{elsayed2018adversarial,yang2021voice2series}, the size of padded learnable parameters is even larger than the original model input. However, in \cite{li2023efficient}, using a larger $l$ did not improve the performance and even caused degradation. When $W$ with a large size is padded on two different input waveforms, $\textbf{\emph{x}}$ and $\textbf{\emph{y}}$, a large portion of the two transformed $\Tilde{\textbf{\emph{x}}}$ and $\Tilde{\textbf{\emph{y}}}$ are identical, which makes them more similar. In reality, an identical waveform indicates a replay, i.e., the same speaker speaking with the same content. Thus, SV models have difficulty distinguishing different speakers in two very similar inputs. In other words, using a large padding is against the purpose of SV. The performance of reprogramming with different sizes of $W$ is evaluated in our experiments in Section \ref{sec:baselines}.


\subsection{Augmentation of input representation}
\label{ssec:adaptationmechanism}
To alleviate the ``identical problem" discussed in the previous section, we propose an augmentation in the padding method of reprogramming.
\cite{wang2020investigation} has shown that random masking input spectrum in training SV models can reduce the risk of overfitting and verification errors. This can be seen as a variation of dropout\cite{srivastava2014dropout} applied to the input level instead of hidden layers. Inspired by \cite{wang2020investigation}, we adopt a similar idea of random masking to alleviate the problem discussed in Section \ref{ssec:limitation}.

The total length of learnable parameters $W$ is denoted as $l$. In each training step, a segment $\Tilde{W}$ with a length of $n=l/k$ is randomly cropped from $W$, where $k$ is a predefined positive integer in the range $[1, l//2]$. The $\Tilde{W}$ is divided into two halves, i.e., $\Tilde{W_L}$ and $\Tilde{W_R}$. Each half is padded to one side of the input waveform as shown in Fig.~\ref{fig:proposeinput}(a). The case of $k=1$ corresponds to the original padding method as in Section \ref{ssec:limitation}.

During inference, the padding method operates as illustrated in Fig.~\ref{fig:proposeinput}(b). $W$ is equally split into $k$ segments, $\{W_{1}, W_{2},..., W_{k}\}$. The single input waveform $x$ replicates into $k$ copies and each copy is padded with one of the segments $W_{i}$ on two sides forming the transformed inputs $\Tilde{X}=\{\Tilde{\textbf{\emph{x}}}_{1},\Tilde{\textbf{\emph{x}}}_{2},..,\Tilde{\textbf{\emph{x}}}_{k}\}$. $k$ embeddings are extracted from $\Tilde{X}$. When comparing $\textbf{\emph{x}}$ with another input speech, $\textbf{\emph{y}}$, the cosine similarity between all pairs of embedding from $\Tilde{X}$ and $\Tilde{Y}$ will be calculated, giving the score matrix $S \in k\times k$. The average value of $S$ is utilized as the criteria to verify $\textbf{\emph{x}}$ and $\textbf{\emph{y}}$. $S_{ij}$ is calculated between $\Tilde{\textbf{\emph{x}}}_{i}$ and $\Tilde{\textbf{\emph{y}}}_{j}$ with different padded parameters, except for the diagonal elements (where $i=j$), which alleviates the ``identical problem" discussed in Section \ref{ssec:limitation}.

We can use a large $l$ of $W$ to ensure the learning ability of the reprogrammed models and increase the diversity of the test waveforms in the evaluation. The padded $\Tilde{W}$ is cropped short in training, which keeps the computation cost low. Moreover, a random crop of $W$ augments the model input during training, decreasing the risk of overfitting.


\begin{table*}[t]
  \caption{Performances on CN-Celeb with different $n$ using different methods. $n$ denotes the number of learnable parameters padded to each speech utterance.}
  \label{tab:combined_performance}
  \centering
  \setlength\tabcolsep{3pt}
  \scalebox{1}{
  \begin{tabular}{lcccccccc}
    \toprule
    \textbf{Model} & \textbf{w/o Adaptation} & \textbf{w Adaptation} & $n=0$ & $n=3,200$ & $n=6,400$ & $n=9,600$ & $n=16,000$ & $n=24,000$ \\
    \midrule
    \midrule
    ECAPA-TDNN-512 & 17.78 & \multirow{3}{*}{\shortstack[l]{$Vanilla$\\$Reprog.$}} & 9.33 & \textbf{8.63} & 9.06 & 9.56 & N/A & N/A \\
    WavLM-Large & 14.62 & & 8.61 & 8.25 & 8.27 & \textbf{8.24} & 8.37 & 8.51 \\
    Wav2Vec2.0-XLSR-53 &  14.23 & & 9.05 & \textbf{8.46} & 9.09 & 9.10 & 9.18 & 9.67 \\
    \midrule
    ECAPA-TDNN-512 & 17.78 & \multirow{3}{*}{\shortstack[l]{$Grad.Est.$\\$Reprog.$}}  & 9.58 & \textbf{9.14} & 9.42 & 9.67 & N/A & N/A \\
    WavLM-Large & 14.62 &  & 8.89 & 8.31 & 8.01 & \textbf{7.83} & 7.83 & 8.10 \\
    Wav2Vec2.0-XLSR-53 & 14.23  &  & 8.53 & 8.35 & 8.32 & \textbf{8.15} & 8.21 & 8.23 \\
    \bottomrule
  \end{tabular}}
  \vspace{-3mm}
\end{table*}

\section{Experimental Settings}
\subsection{Network structures}
\subsubsection{Embedding extraction models}
\label{sssec:backbone}
Three pre-trained models are utilized in this work: ECAPA-TDNN-512\cite{desplanques20_interspeech}, WavLM-Large\cite{chen2022wavlm}, and Wav2Vec2.0-XLSR-53\cite{conneau21_interspeech,baevski2020wav2vec}. They are popular deep networks for speaker embedding extraction with state-of-the-art performance. The dimension of extracted speaker embedding is $256$ for all models.

The ECAPA-TDNN-512 model is pre-trained by ourselves on the development set of VoxCeleb2\cite{nagrani2020voxceleb}, achieving the Equal Error Rate (EER) of $1.12\%$ on the evaluation set of VoxCeleb1\cite{nagrani2017voxceleb}(Vox1.O). The number of parameters in this model is around 5.95 million ($M$).

WavLM and Wav2Vec2.0 are two large-scale self-supervised networks that were successfully adopted in the SV task \cite{chen2022large,DBLP:conf/interspeech/Chen0000WL00YW22}. The number of parameters in these two models is 316.62$M$ and 315$M$, respectively. We retrieve their pre-trained models directly from \url{https://github.com/microsoft/UniSpeech/tree/main/downstreams/speaker_verification} without making any modification. They achieve the EERs of $0.431\%$ and $0.564\%$ on Vox1.O, respectively. 

\subsubsection{Adaptation models}
\label{ssec:adaptationmodels}
In accordance with the experimental setup delineated in \cite{li2023efficient}, we architect our adaptation model with some minor modifications. The gradient estimation network used in \cite{li2023efficient} is a lightweight ECAPA-TDNN-512\cite{desplanques20_interspeech}, which is a convolution-based network. Convolution operations are adept at processing adjacent feature neighborhoods, while their capacity for global information integration is inherently restricted. To better consolidate the global information for verification, three self-attention\cite{NIPS2017_3f5ee243} blocks are added preceding each Res2Net block. The three self-attention blocks share weights to decrease the total number of additional parameters.
The channel $C$ of the gradient estimation network in \cite{li2023efficient} is set to $32$. This type of reprogramming is denoted as gradient estimated reprogramming ($Grad.Est.Reprog.$), in contrast to vanilla reprogramming ($Vanilla\ Reprog.$). 

The two-layer fully-connected layers at the output stage of the fixed pre-trained backbone \cite{li2023efficient} are replaced with one single linear projection layer for better classification and simplicity.

\subsection{Datasets}
The embedding extraction models described in Section~\ref{sssec:backbone} are pre-trained or fine-tuned from VoxCeleb, where most of the speech utterances are in English\cite{nagrani2020voxceleb}. To perform the cross-language SV adaptation and evaluation, CN-Celeb1\cite{fan2020cn} is utilized, which is a challenging Chinese dataset. Domain adaptation is performed using the training set of CN-Celeb1 and evaluated on the evaluation set of it. The training and test set contain $800$ and $200$ speakers, respectively.

\subsection{Training settings}
For training, the batch size is set to $128$. In each training step, a 2-second segment is randomly cropped from each utterance. The raw waveforms of the cropped segments are reprogrammed based on Section~\ref{ssec:adaptationmodels} and transformed into 64-dimension log Mel-filterbanks (FBank) as model inputs.
The learnable parameters $W$ are initialized following a Gaussian distribution. The adaptation model is trained by the Adam optimizer with a weight decay of $1e-4$. The learning rate is initialized as $1e-3$ and reduced by a ratio of $10$ at the $10^{th}$, $15^{th}$ epoch, respectively. The total training epoch is set to $20$. AAM-Softmax\cite{deng2019arcface} is utilized as the loss function during training.


\begin{table*}[t]
  \caption{Performances of models on the test set of CN-Celeb. $n$ denotes the number of learnable parameters appended to each copy of a speech utterance. $l$ denotes the total number of parameters padded. $k=l/n$ is the number of replicated copies of speech utterances. Above the horizontal rules are the results of the raw padding method, i.e., $k=1$, while below them corresponds to the augmented padding method, i.e., $k>1$.}
  \vspace{1mm}
  \label{tab:result}
  \centering
  \setlength\tabcolsep{3pt}
  \scalebox{1}{
   \begin{tabular}{lcccccccccccc}
    \toprule
      & \multicolumn{4}{c}{\textbf{ECAPA-TDNN-512}} & \multicolumn{4}{c}{\textbf{WavLM-Large}} & \multicolumn{4}{c}{\textbf{Wav2Vec2.0-XLSR-53}} \\
     \cmidrule(lr){2-5} \cmidrule(lr){6-9} \cmidrule(lr){10-13}
    \textbf{Adaptation}
    & \textbf{EER ($\%$)} & \textbf{$n$ ($\times 10^{3}$)} & \textbf{$l$ ($\times 10^{3}$)} & \textbf{$k$}
    & \textbf{EER ($\%$)} & \textbf{$n$ ($\times 10^{3}$)} & \textbf{$l$ ($\times 10^{3}$)} & \textbf{$k$}
    & \textbf{EER ($\%$)} & \textbf{$n$ ($\times 10^{3}$)} & \textbf{$l$ ($\times 10^{3}$)} & \textbf{$k$} \\
    \midrule
    \midrule
    \multirow{6}{*}{\shortstack[l]{$Vanilla$\\$Reprog.$}}
    & \textbf{8.63} & \textbf{3.2}& 3.2 & 1 & 8.25 & 3.2 & 3.2 & 1 & 8.46 & 3.2 & 3.2 & 1 \\
    & 9.06 & 6.4 & 6.4 & 1 & 8.27 & 6.4 & 6.4 & 1 & 9.09 & 6.4 & 6.4 & 1 \\
    & 9.56 & 9.6 & 9.6 & 1 & 8.24 & 9.6 & 9.6 & 1 & 9.1 & 9.6 & 9.6 & 1 \\
     \cmidrule(lr){2-5} \cmidrule(lr){6-9} \cmidrule(lr){10-13} 
    & 9.11 & 2.4 & 4.8 & 2 & 8.35 & 3.2 & 6.4 & 2 & \textbf{8.31} & \textbf{3.2} & 6.4 & 2 \\
    & 9.19 & 3.2 & 6.4 & 2 & \textbf{8.24} & \textbf{6.4} & 12.8 & 2 & 8.64 & 6.4 & 12.8 & 2 \\
    & 9.15 & 3.2 & 9.6 & 3 & 8.27 & 9.6 & 19.2 & 2 & 8.91 & 9.6 & 19.2 & 2 \\
    \midrule
    \multirow{6}{*}{\shortstack[l]{$Grad.Est.$\\$Reprog.$}}
    & 9.14 & 3.2 & 3.2 & 1 & 8.31 & 3.2 & 3.2 & 1 & 8.35 & 3.2 & 3.2 & 1 \\
    & 9.42 & 6.4 & 6.4 & 1 & 8.01 & 6.4 & 6.4 & 1 & 8.32 & 6.4 & 6.4 & 1 \\
    & 9.67 & 9.6 & 9.6 & 1 & \textbf{7.83} & \textbf{9.6} & 9.6 & 1 & \textbf{8.15} & \textbf{9.6} & 9.6 & 1 \\
     \cmidrule(lr){2-5} \cmidrule(lr){6-9} \cmidrule(lr){10-13}
    & 9.43 & 2.4 & 4.8 & 2 & 7.94 & 3.2 & 6.4 & 2 & 8.39 & 3.2 & 6.4 & 2 \\
    & \textbf{9.06} & \textbf{3.2} & 6.4 & 2 & 7.97 & 6.4 & 12.8 & 2 & 8.46 & 6.4 & 12.8 & 2 \\
    & 9.12 & 3.2 & 9.6 & 3 & 7.84 & 9.6 & 19.2 & 2 & 8.40 & 9.6 & 19.2 & 2 \\
    \bottomrule
  \end{tabular}}
  \vspace{-3mm}
\end{table*}

\begin{table}[t]
  \caption{Performances on CN-Celeb with different training data sizes.
  }
  \vspace{0.5mm}
  \label{tab:random}
  \centering
  \setlength\tabcolsep{3pt}
  \scalebox{1}{
   \begin{tabular}{llcccc}
    \toprule
      \multirow{2}{*}{\shortstack[l]{\textbf{Model}}} & \multirow{2}{*}{\shortstack[l]{\textbf{Adaptation}}} & \multirow{2}{*}{\shortstack[c]{\textbf{No. of}\\ \textbf{speakers}}} & \multirow{2}{*}{$n=0$} & \multicolumn{2}{c}{$n=3,200$} \\
      \cmidrule(lr){5-6}
      & & & & $k=1$ & $k=2$ \\
    
    \midrule
    \midrule
    \multirow{6}{*}{\shortstack[l]{ECAPA-\\TDNN-\\512}} & \multirow{3}{*}{\shortstack[l]{$Vanilla$\\$Reprog.$}}
    & 20 & 12.52 & 12.57 & \textbf{12.45} \\
    & & 50 & 12.21 & 12.03 & \textbf{11.99} \\
    & & 100 & 12.21 & 11.96 & \textbf{11.77} \\
    \cmidrule{2-6}
    & \multirow{3}{*}{\shortstack[l]{$Grad.Est.$\\$Reprog.$}}
    & 20 & 12.50 & \textbf{12.48} & 12.59 \\
    & & 50 & 12.18 & \textbf{12.13} & 12.17 \\
    & & 100 & 12.17 & 12.13 & \textbf{12.00} \\
    \midrule
    \multirow{6}{*}{\shortstack[l]{WavLM-\\Large}} & \multirow{3}{*}{\shortstack[l]{$Vanilla$\\$Reprog.$}}
    & 20 & 11.15 & \textbf{11.09} & \textbf{11.09} \\
    & & 50 & 10.92 & 10.88 & \textbf{10.87} \\
    & & 100 & \textbf{10.75} & 10.81 & 10.83 \\
    \cmidrule{2-6}
    & \multirow{3}{*}{\shortstack[l]{$Grad.Est.$\\$Reprog.$}}
    & 20 & 11.14 & \textbf{10.93} & \textbf{10.93} \\
    & & 50 & 10.91 & \textbf{10.78} & 10.79 \\
    & & 100 & 10.81 & 10.80 & \textbf{10.67} \\
    \bottomrule
  \end{tabular}}
  \vspace{-3.5mm}
\end{table}

\section{Results and Analysis}
\label{sec:results}
Different numbers of learnable parameters and padded lengths per segment are evaluated in our experiments. The padded length ranges from $0.2s$ to $1.5s$ duration of a waveform, i.e., the value of $n$ being $3,200$ to $24,000$ for $16k$ Hz input audio. 


\subsection{Baselines}
\label{sec:baselines}

\begin{figure}[ht]
  \centering
  \scalebox{0.8}{
  \includegraphics[width=\linewidth]{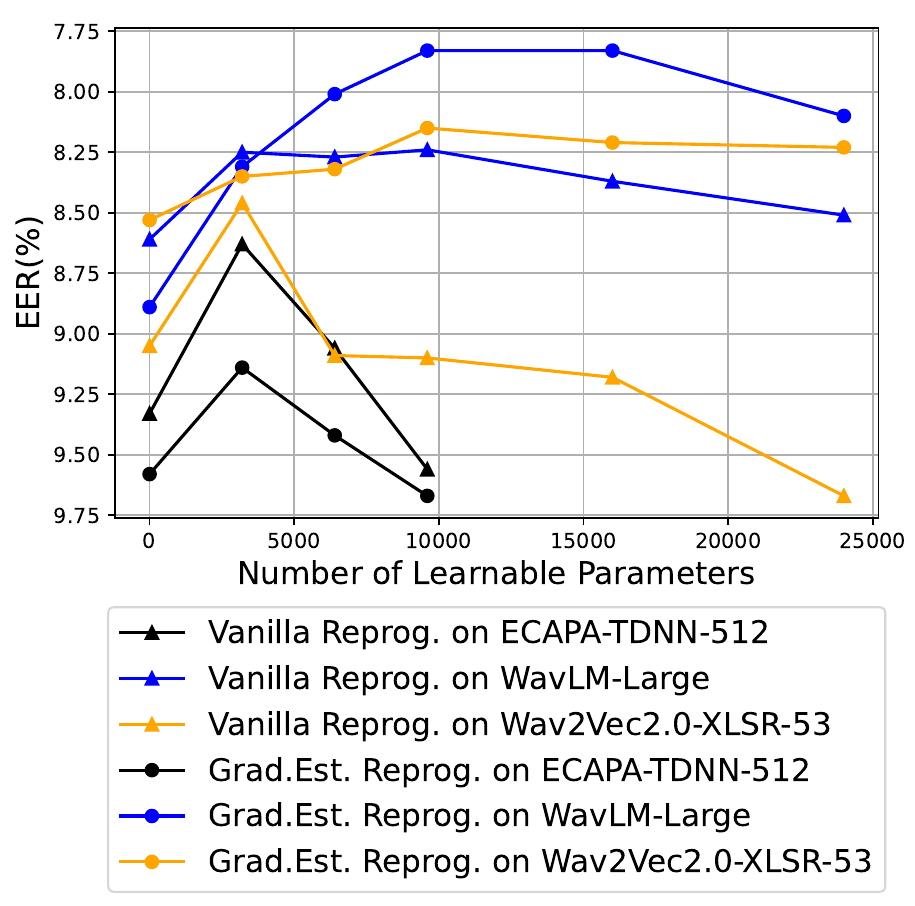}
  }
  \caption{Performance of reprogrammed models (EER) v.s. number of padded learnable parameters ($n$)}
\label{fig:scatter}
\vspace{-1mm}
\end{figure}

The results of reprogramming with the raw padding method on the three pre-trained models are shown in Table \ref{tab:combined_performance} and Fig.~\ref{fig:scatter}. The results present the relationship between the adaptation model's performance, which is measured by EER, and the number of padded parameters $n$. An initial decrease in EER is observed as $n$ increasing from $0$ to $3,200$, indicating the contribution of model reprogramming. $n=0$ refers to the case where no learnable parameters are padded, and only the SV backend is trained during training.

However, as discussed in Section \ref{ssec:limitation}, this performance enhancement does not persist as $n$ continues to rise beyond $3,200$ to $9,600$. A critical threshold is encountered, beyond which the EER starts to increase. For ECAPA-TDNN-512, The vanilla reprogramming and gradient estimated reprogramming both have the peak performance at $n=3,200$, which subsequently deteriorates as $n$ continues to increase.

WavLM-Large and Wav2Vec2.0-XLSR-53 are much larger than ECAPA-TDNN-512. Additionally, they are pre-trained with much more data during self-supervised learning, which gives them more robust learning and generalization abilities. They outperform ECAPA-TDNN-512 when using the same padding length $n$, and they can process a larger number of padded parameters without performance degradation. However, they also demonstrate the same phenomenon when $n$ keeps increasing. As the value of $n$ exceeds a specific value, e.g., $n=9,600$ for Wav2Vec2.0-XLSR-53 and $n=16,000$ for WavLM-Large, the performance of reprogrammed adaptation models starts to decrease. 


\subsection{Results on the augmented padding}
The augmented input representation method is evaluated, and the results are shown in Table \ref{tab:result} in comparison with the raw padding method. The EER is used to measure the performance of the adaptation model, and different settings of $n$, $l$, and $k$ defined in Section \ref{ssec:adaptationmechanism} are evaluated on their effects on the performance.

Improvements can be observed among the three different pre-trained models. For $Grad.Est.\ Reprog.$ on ECAPA-TDNN-512, utilizing the augmented input with $n=3,200$, $l=6,400$, and $k=2$ can decrease the EER compared to cases where $l=3,200$ or $l=6,400$ but $k=1$. Introducing the augmentation to $Vanilla\ Reprog.$ on WavLM-Large and Wav2Vec2.0-XLSR-53 also demonstrates the effectiveness where lower EER cannot be achieved with larger $n$, i.e., the number of parameters padded during training.

The improvement is not consistent across all different settings. Augmentation brings better performance than $k=1$, given the same value of $n$ or $l$, but it does not give the optimal performance under the specific range of numbers of learnable parameters introduced. From the trend shown in the table, the performance is primarily brought by the more advanced model architecture or larger model scale.

The augmented input representation is considered to alleviate the ``identical problem" brought by a larger portion of the same content, but the results show that the improvement is limited despite the reduced parameters during training. It is indicated that the performance is still limited by $l$, i.e., the total parameters introduced. Learning ability is primarily determined by the capacity of the pre-trained model, which can handle a fixed range of learnable parameters for model reprogramming.




\subsection{Results on different scales of data}
Domain adaptation with a limited quantity of data is challenging but common in practical applications. The effectiveness of the proposed method with different scales of data is evaluated on ECAPA-TDNN-512 and WavLM-Large in this section. The datasets are constructed in three scales: $20$, $50$, and $100$ speakers. The speakers are randomly sampled from CN-Celeb 1. A minimum of $20$ and a maximum of $50$ utterances are randomly selected from each speaker. During training, the model is adapted to the constructed dataset for $100$ epochs, and the learning rate is reduced by a ratio of $10$ at the $60^{th}$ and $80^{th}$ epoch. The fully-connected layer at the output stage of the fixed pre-trained models is dropped to reduce overfitting and achieve higher efficiency. A fixed setting of $n=3,200$ is utilized in the experiments. Each scale of the dataset is randomly generated five times, and the average results are reported in Table ~\ref{tab:random}.

The results show that the performance for all four settings increases as the training data sizes increase. Introducing learnable parameters leads to a decrease in EER. As the results have shown, the proposed augmentation can achieve comparable or surpassing performance as compared to the raw padding method across different models and settings. 

Despite the removal of the fully-connected layer, the adaptation models are more inclined to overfitting when available training data is limited \cite{Ying_2019}. The results indicate that the overfitting is alleviated in adaptation model training when the augmented input representation is applied, which can be useful in practical resource-constrained scenarios.

\subsection{Limitations and future work}
The experimental results demonstrate that reprogramming can enhance cross-language adaptation in SV systems, but certain limitations remain. We focus primarily on model reprogramming without comparing it with the fully finetuning method and other advanced model tuning methods like adapter or prompt tuning. Although sharing some similarities with prompt tuning, reprogramming focuses on language adaptation, while prompt tuning is generally utilized in task adaptation \cite{pmlr-v202-melnyk23a}. Nevertheless, future work can explore the effectiveness of model reprogramming compared to other advanced adaptation methods or the fully finetuning method. The problem found when the padded length becomes huge may be due to the potential overfitting since the data size remains unchanged. Future work can explore better ways of input transformation for SV.

\section{Conclusions}
In this work, we revisit the use of the reprogramming method for cross-language domain adaptation in SV and investigate its limitations. The experiments demonstrate that reprogramming is an effective adaptation method in different scales of SV models and datasets. However, the performance of reprogrammed outputs is mainly determined by the pre-trained models, and increasing the padded learnable parameters does not contribute to the improvement after a criteria value. Larger scales of SV models have more powerful capacities to process longer padded parameters and give better results.

\bibliographystyle{IEEEtran}

\bibliography{main}


\end{document}